%% file: 20mALPS.tex
\documentclass[10pt,letterpaper]{article}
\usepackage{opex3}
\usepackage{color}
\usepackage{pstool}
\usepackage{epstopdf}
\begin{document}

\title{Characterization of optical systems for the ALPS\,II experiment}

\author{Aaron D. Spector$^{1,+}$, Jan H. P\~old$^{2,*}$, Robin B\"ahre$^{3,4}$, Axel Lindner$^{2}$ and Benno Willke$^{3,4}$}

\address{$^1$Institut f\"ur Experimentalphysik, Universit\"at Hamburg, Hamburg, Germany.\\
$^2$Deutsches Elektronen-Synchrotron (DESY), Hamburg, Germany.\\
$^3$Max Planck Institute for Gravitational Physics (Albert Einstein Institute), Hannover, Germany.\\
$^4$Institut f\"ur Gravitationsphysik der Leibniz Universit\"at Hannover, Hannover, Germany}

\email{$^+$aaron.spector@desy.de}
\email{$^*$jan.pold@desy.de} 



\begin{abstract}
ALPS\,II is a light shining through a wall style experiment that will use the principle of resonant enhancement to boost the conversion and reconversion probabilities of photons to relativistic WISPs. This will require the use of long baseline low-loss optical cavities. Very high power build up factors in the cavities must be achieved in order to reach the design sensitivity of ALPS\,II. This necessitates a number of different sophisticated optical and control systems to maintain the resonance and ensure maximal coupling between the laser and the cavity. In this paper we report on the results of the characterization of these optical systems with a 20\,m cavity and discuss the results in the context of ALPS\,II.
\end{abstract}


\bibliographystyle{unsrt}
\bibliography{paper}

\section{Introduction}

Long baseline, low-loss optical cavities are used in a number of different applications including gravitational wave detectors \cite{ligodet} and magnetic vacuum birefringence experiments \cite{pvlas}. In addition to being essential to these projects, optical cavities can also be used to improve the sensitivity of Light-Shining-through-Wall (LSW) experiments \cite{rc,Fukuda,Sikivie}.

These experiments are designed to detect theoretically proposed and astrophysically motivated weakly-interacting-sub-eV-particles (WISP) \cite{Olive}. This experimental approach is characterized by shining a light source at an opaque wall. Before the wall some photons will convert to WISPs and these WISPs will traverse the wall into the regeneration side of the experiment \cite{Graham}. There, the WISPs will reconvert to photons which are measured with a single photon detection scheme. Optical cavities have been used before the wall to increase the number of photons creating WISPs, such as in the Any Light Particle Search I (ALPS\,I). A cavity can also be used after the wall to increase the probability that WISPs will reconvert to a photons. The cavities before and after the wall will be referred to as the production cavity and the regeneration cavity, respectively.  LSW experiments typically include dipole magnets in their design to search for axion-like particles.

ALPS\,II \cite{alpstdr} is a second generation LSW experiment currently being prepared at DESY in Hamburg. It will consist of a production and a regeneration cavity each with a length of 100\,m. Each of the cavity Eigenmodes will pass through a string of 5 T superconducting dipoles that form 88\,m of magnetic length. The design sensitivity of ALPS\,II will surpass current limits for the detection of axion-like particles by three orders of magnitude set by other experiments of its kind \cite{ALPSI,OSQAR,Betz}.

The cavities on both sides of the wall must maintain their resonance condition for an extended period of time to ensure that the experiment meets its sensitivity goals. This requires a number of different controlled optical systems. The optical technologies and the frequency and alignment control schemes for ALPS\,II were tested using an optical cavity with a length of 20\,m. Furthermore, an in-depth characterization of the optical losses, control loop couplings, and environmental noise sources are required to understand and probe the stability and robustness of the setup.

In this article we report on the progress of these efforts. Section \ref{sec:SET} describes the optical setup of this test experiment. The experimental results are presented and discussed in Section \ref{sec:Meas} and conclusions are drawn in Section \ref{sec:CON}.

\section{Setup}
\label{sec:SET} 
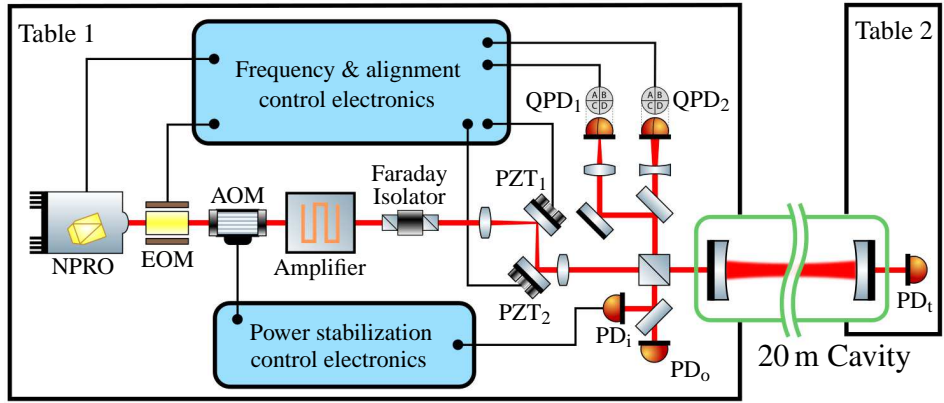
\begin{figure}[t] 
  \centering
  \input{figures/fullFINALsmall.tex} 
  \caption{\label{fig:SET} Diagram of the current optical setup with control electronics for the frequency stabilization, automatic alignment, and power stabilization. An NPRO laser provides the beam which is modulated  with EOM and AOM, and then amplified to an output power of 35W. A fraction of this light is incident on the 20\,m cavity. The photodetectors QPD$_1$ and QPD$_2$ in reflection of the cavity are used to generate the PDH and automatic alignment error signals. The PDH error signal is fed back to a PZT on the NPRO crystal as well as a heater that actuates the crystal temperature. The automatic alignment error signals are sent to the alignment PZTs which are located at the same Gouy phase position as their corresponding photodetectors. This is displayed by PZT$_1$ and QPD$_1$ both occupying a waist position while PZT$_2$ and QPD$_2$ are both located at a far field position in the beam. A power stabilization scheme uses the AOM as an actuator with the sensing performed by PD$_{\rm i}$. PD$_{\rm o}$  measures the out-of-loop power noise. The cavity end mirrors are located on separate tables and the vacuum system is shown in green. PD$_{\rm t}$ measures the power noise in transmission of the cavity.}
\end{figure}

The setup consists of a 20\,m optical cavity with the mirrors on two separate optical tables as shown in Figure \ref{fig:SET}. The end mirrors are two inches in diameter with a radius of curvature of $23.2\pm0.5$\,m giving the cavity a nearly confocal configuration. The transmissivity of the mirrors was measured to be $878\pm2$\,ppm. The resonant Eigenmode of the cavity that is used for stabilizing the laser frequency is the Hermite-Gauss $U_{00}$ mode defined in \cite{Kogelnik}. It has a beam waist of $1.97\pm0.02$\,mm and a mode radius of $2.62\pm0.01$\,mm at the end mirrors. For a loss free cavity the calculated Finesse is $3577\pm8$ and the linewidth is $2097\pm5$\,Hz. This corresponds to a storage time of $151.8\pm0.3$\,$\mu$s.

A vacuum system houses the two cavity mirrors and the measurements were taken at a pressure of 5$\cdot$10$^{-5}$\,mbar. A dedicated lab hosts the 20\,m vacuum envelope and the optical tables in a clean environment.

 The laser system consists of a non-planar-ring-oscillator (NPRO) master laser and a Nd:YVO$_4$ amplifier \cite{Frede:07}. To avoid thermal effects in this experiment we inject only 50\,mW of power into the cavity. The laser, developed for the LIGO gravitational wave detector, delivers a single frequency beam with a high fundamental mode content.

The laser frequency is stabilized to a resonance of the cavity using the Pound Drever Hall (PDH) frequency stabilization technique \cite{Drever,Black}. The required phase modulation at 2.5\,MHz is created with an electro optic modulator (EOM). For frequency actuation the control loop uses a Piezo (PZT) actuator mounted to the NPRO crystal for fast frequency correction and the temperature of the NPRO crystal to stabilize long term frequency drifts. The control loop operates with unity gain frequencies up to 55\,kHz.

In addition, an automatic alignment system was built using the quadrant photodetectors QPD$_1$ and QPD$_2$ to do differential wavefront sensing in order to maintain the alignment of the injected beam with respect to the cavity Eigenmode \cite{AA}. The actuation is done with the two axes alignment piezo mirrors PZT$_1$ and PZT$_2$ (see Figure \ref{fig:SET}). We were able to realize unity gain frequencies of 170 Hz actuating on PZT$_2$ and  unity gain frequencies of 15 Hz actuating on PZT$_1$. 

The photodetector PD$_{\rm i}$ is used as a sensor for the power stabilization while the Acoustic Optic Modulator (AOM) is used as an actuator for the control loop. The out-of-loop photodetector PD$_\mathrm{o}$ measures the resulting power noise at a pick off in front of the cavity. We achieved a unity gain frequency of 1 kHz with a suppression of 60\,dB at 10\,Hz. The photodetector PD$_\mathrm{t}$ on table 2 measures the light transmitted through the cavity.

\section{Measurements}
\label{sec:Meas}

ALPS\,II aims to achieve high power buildup factors of the optical cavities as well as 95\% spatial overlap between their Eigenmodes and the mode of the laser. Therefore, the characterization and mitigation of losses in the cavity and a high visibility is essential.

The visibility of the cavity, which refers to the ratio of power reflected from the incoupling mirror on and off resonance, was found to be greater than 95\%. This indicates that the higher order spatial mode content and mode mismatch of the input beam is less than 5\%.


Two independent methods are used to accurately measure the internal cavity losses. The first method measures the pole frequency of the cavities low-pass filter property \cite{Mueller}. The linewidth is equal to twice the pole frequency and a high finesse corresponds to a high power build up factor in the cavity. This measurement reveals a linewidth of $2460\pm20$\,Hz in vacuum and $2370\pm70$\,Hz in air which corresponds to a finesse of $3050\pm20$ and $3170\pm90$, respectively. 

The second method measured the cavity storage time by observing the exponential decay of the light leaking from the output mirror of the cavity after the input power is instantaneously reduced \cite{cavloss}. The cavity storage time ($\tau_{\rm{storage}}$) is $134\pm10$\,$\mu$s for the in air case leading to a finesse of $3160\pm240$.
 Thus, the two results agree very well. 
 
 This corresponds to a measurement of $230\pm50$\,ppm internal losses within the cavity for the pole frequency measurement in air. The pole frequency measurement in vacuum yields internal losses of $300\pm20$\,ppm. 
The additional losses could potentially be explained by scattering on point defects, dust, or the micro-roughness of the mirror surfaces. The intracavity losses appear to have some dependence on the location of the Eigenmode on the end mirrors. 
Further investigations of this are required.

The frequency stabilization scheme is very robust. Long term drifts are compensated by the automatic alignment and the feedback to the temperature of the NPRO. The range for the latter is up to 30\,GHz corresponding to cavity length changes of 2\,mm. A stable resonance condition was maintained for more than 48\,hours before the control loop was manually disengaged.

The optical systems for the current setup are evaluated by measuring their ability to suppress the relative power noise in transmission of the cavity.  Suppressing this noise while minimizing the amount of light reflected from the input mirror ensures that there is maximum power circulating inside the cavity. We investigated three noise sources which contribute to the cavity transmission power noise. These are the differential frequency noise between the laser and the cavity resonance, the alignment noise between the input beam and the cavity Eigenmode, and the power noise of the laser. The following sections detail measurements of all three noise sources and project them in terms of their contribution to the relative power noise in transmission of the cavity.

\subsection{Cavity Frequency Noise}
\label{sec:FN} 

\begin{figure}[t]
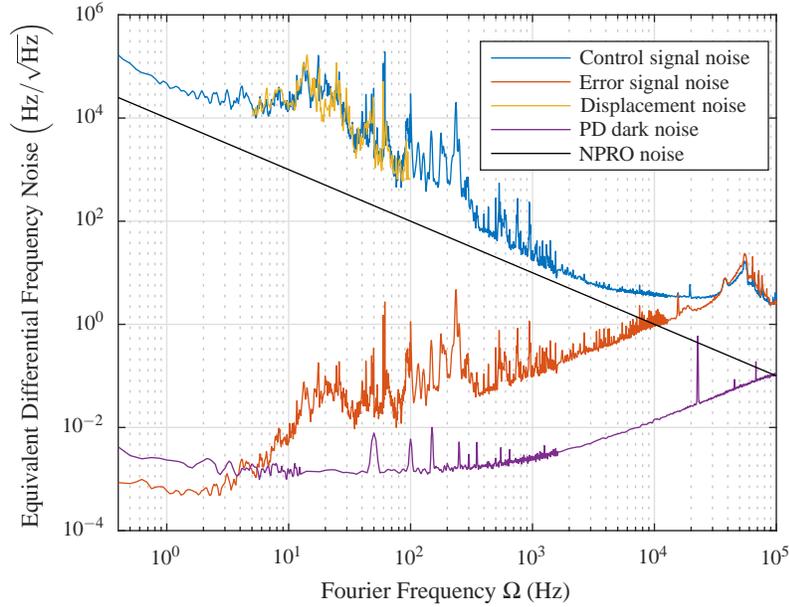
 
    \centering
  \psfragfig{figures/FNb}
   \caption{\label{fig:FN} Spectral densities of the control and error signals of the frequency stabilization control loop calibrated in terms of differential frequency noise are shown in blue and orange respectively. A non-coincident geophone measurement of the differential displacement noise between the optical tables is plotted in yellow. This measurement represents the length noise of the cavity due to ground motion. The electronic noise of the PDH sensing is shown in purple and a typical trace of the NPRO frequency noise is plotted in black \cite{NPROcharac}.}
\end{figure}

Figure \ref{fig:FN} shows spectra of the electronic noise at various points of the frequency control electronics. These spectra are calibrated as equivalent differential frequency noise which refers to the difference between the resonant frequency of the cavity and the frequency of the laser. We will refer to this as the differential frequency. The blue curve is the control signal that is sent to the laser piezo to stabilize its frequency with respect to the cavity. When expressed as frequency noise below the unity gain frequency of the control system this gives the differential frequency fluctuation of the uncontrolled system (free running).

The orange curve is the spectrum of the noise at the error point of the control electronics divided by the cavities transfer function. This division is required to calibrate the error point noise in terms of differential frequency noise. Therefore the orange curve represents an in-loop measurement of the differential frequency noise with suppression from the frequency stabilization control scheme. The system was able to reach a unity gain frequency up to 55\,kHz before becoming unstable. At 20\,Hz we have roughly six orders of magnitude suppression of the free running noise. 

The yellow curve is a measurement of the differential displacement noise from two Mark Products L-4C geophones placed on tables 1 and 2 and oriented in the direction of the optical axis of the cavity. The differential displacement noise measured by the geophones is calibrated to differential frequency noise in the cavity with Equation \ref{equ:SEI}.
\begin{equation}
\label{equ:SEI}
\frac{\Delta f}{f}=\frac{\Delta L}{L}
\end{equation}
Here $\Delta f$ is the frequency change of the cavity resonance due a displacement $\Delta L$ between the mirrors of the cavity. $f$ is the laser frequency and $L$ distance between the cavity mirrors. 

Figure \ref{fig:FN} shows that the control signal noise closely follows the seismic noise from 5\,Hz to 100\,Hz. This demonstrates that in this region the displacement noise between the mirrors couples to the differential frequency noise. Differences between these two curves can be attributed to the fact that these measurements were not performed coincidentally. Differential displacement noise dominates the RMS for frequencies below 400\,Hz. 

The black line is a representation of the typical frequency noise of a free running NPRO laser like the one we are using \cite{NPROcharac}. As Figure \ref{fig:FN} shows, NPRO noise is the dominating noise source from 400\,Hz to 10\,kHz. Above 10\,kHz the control signal noise is elevated by the limited phase margin of the control electronics. 

 
 The purple curve is a measurement of the error signal noise divided by the cavity transfer function when the laser system is turned off. This shows the effect of photodetector dark noise on the frequency stabilization scheme and represents a lower limit on the possible suppression of the frequency noise. 
 
The differential frequency noise can be projected to power noise in transmission of the cavity. For this projection we model the remaining phase noise as if it were caused by the laser. This is done by solving the Equation \ref{equ:CAV} with the error signal representing the differential phase $\phi(t)$ \cite{Benno}.
\begin{equation}
\label{equ:CAV}
E_{\rm cav}(t) = \sqrt{R_1 R_2(1-L)}E_{\rm cav}(t-\tau)+\sqrt{T_1}E_{\rm in}(t)\cdot e^{-i\phi(t)}
\end{equation}
Here $E_{\rm cav}(t)$ gives the field circulating inside the cavity. It is clearly dependent on phase and amplitude of the input field $E_{\rm in}(t)$. There is no phase term in $E_{\rm cav}(t-\tau)$ since the cavity is assumed to be on resonance. $R_1$ and $R_2$ are the power reflectivities of the in and out coupling mirrors, while $T_1$ is the power transmissivity of the in coupling mirror. $\tau$ is the light travel time for one round trip through the cavity and $L$ represents the extra losses inside the cavity. This equation was solved iteratively using a time series of the error signal noise calibrated in terms of phase. 

From this the power in transmission of the cavity can be simulated. The result of this projection can be seen in Figure \ref{fig:RIN} and is discussed in Section \ref{sec:RIN}. It is important to note that a small offset in the error signal will increase the coupling from differential frequency noise to power noise in transmission of the cavity. Therefore we minimized the linear component of the coupling before the measurement. 

\subsection{Pointing noise}
\label{sec:PN}

Small misalignments between the Eigenmode of the cavity and the input beam can be expressed as the latter having a small amount of power in the Hermite-Gauss $U_{01}$ and $U_{10}$ modes defined by the cavities Eigenmode basis. Equation \ref{equ:EP} shows how this can be quantified in terms of the parameter $\epsilon$ by writing the input laser as a superposition of the spatial eigenmodes of the cavity \cite{AA}.
\begin{equation}
\label{equ:EP}
U_{\rm in} \approx A[U_{00}+\epsilon_x U_{01}+\epsilon_y U_{10}]
\end{equation}
Here $\epsilon$ represents the relative field amplitude of the $U_{01}$ or $U_{10}$ modes with $U_{01}$ corresponding to a horizontal misalignment and the $U_{10}$ to a vertical misalignment. In addition to having vertical and horizontal components, $\epsilon$ also has real and imaginary parts as shown in Equation \ref{equ:EPb} for the $x$ component.
\begin{equation}
\label{equ:EPb}
\epsilon_x = \frac{\delta \hspace{-0.015cm}x}{w_o}+i\frac{\delta \hspace{-0.015cm}\phi_x}{\theta_o}
\end{equation}
The real part of $\epsilon_x$ describes the relative lateral displacement between the waist positions of the modes. $\delta \hspace{-0.015cm}x$ is the lateral offset between the parallel optical axes of the modes and $w_o$ is the beam radius at the waist. The imaginary part of $\epsilon_x$ gives the angular misalignment of the two beams. $\delta \hspace{-0.015cm}\phi_x$ is the angular mismatch between the optical axes while $\theta_o$ is the divergence half angle of the beams. 

To simulate the relative pointing noise in terms of power noise in transmission of the cavity, the power of the input beam in the $U_{00}$ mode of the cavity is the important parameter. For values of $|\epsilon|\ll1$ the spectra of $|\epsilon|^2$ gives the contribution of the alignment noise to the transmission power noise. This spectrum was found by measuring the time series of the four automatic-alignment error signals, calibrating them in terms of $\epsilon$, assuming a perfect decoupling between the channels, and summing their squares. The calibrated  spectrum was then filtered by the transfer function of the cavity to give the final projection in terms of power noise in transmission of the cavity and the result is shown in Figure \ref{fig:RIN}. 

\subsection{Power noise}
\label{sec:LIN} 

The laser power noise spectrum is measured with the out-of-loop photodetector PD$_{\rm o}$. The performance is currently limited by the electronic noise of the in-loop photodetector below a Fourier frequency of 500\,Hz. However, peaks in the power noise spectrum which dominate the free running RMS noise are suppressed by the control loop.
The contribution of the intrinsic power noise of the input beam to the power noise in transmission of the cavity can be found by multiplying the laser's power noise spectral density by the transfer function of the cavity as shown in Figure \ref{fig:RIN}. 

\subsection{Relative power noise in transmission of the cavity}
\label{sec:RIN}

\begin{figure}[t]
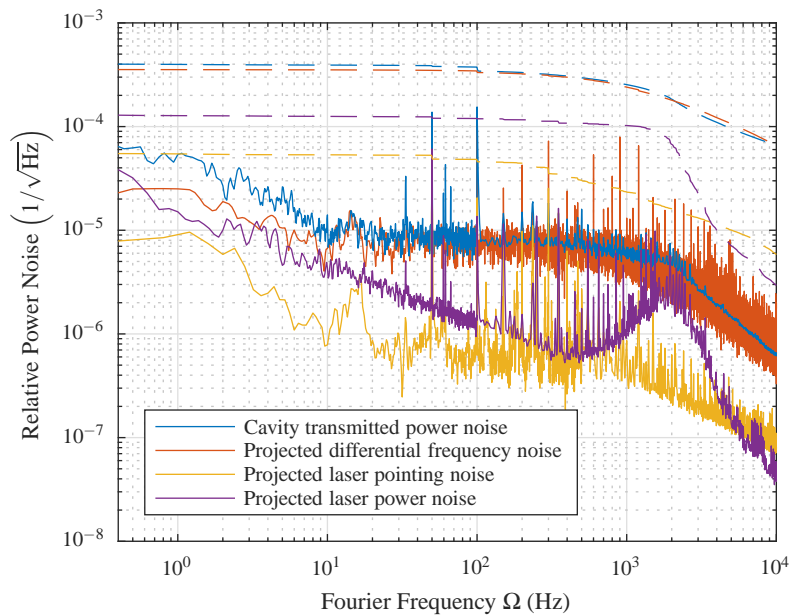
 
  \centering
  \psfragfig{figures/RINhf}
   \caption{\label{fig:RIN} Noncoincidently measured linear spectral densities of the relative power noise in transmission of the cavity along with projections of the transmitted power noise due to differential frequency, pointing, and input beam power noise. The dashed lines denote the integrated RMS of each of the spectral densities.}
\end{figure}


Figure \ref{fig:RIN} shows a plot of the relative power noise in transmission of the cavity along with projections of the three noise sources we investigated. These are the differential frequency and pointing noise of the input beam with respect to the cavity Eigenmode and the power noise of the input beam. The total RMS of the measured transmission power noise spectrum is $4.0\times10^{-4}$. The three projections have a total RMS of $3.5\times10^{-4}$, $5\times10^{-5}$, and $1.3\times10^{-4}$, for the projections of the differential frequency noise, pointing noise, and input beam power noise respectively. The RMS of sum of the three noise projections is $3.8\times10^{-4}$. Thus we can attribute 95\% of the RMS of the cavity transmission power noise to these three noise sources. 


The spectrum of the power noise in transmission of the cavity has three regions with distinct features. A low frequency region below 20\,Hz where the noise increases as the frequency decreases, a middle region from 20\,Hz to 2\,kHz in which the spectrum is relatively flat, and a high frequency region above 2\,kHz at which point the noise rolls off with a $f^{-1}$ slope. 
The broadband noise below 20\,Hz is responsible for 3\% of the RMS and the peaks at 100\,Hz and 50\,Hz are responsible for 8\% and 2\% of the RMS respectively. The noise above 100\,Hz accounts for 85\% of the RMS.

The projected differential frequency noise accounts for 90\% of the RMS power noise in transmission of the cavity. The spectrum of this projection closely follows the shape of the measured transmission power noise. A broad peak at 15\,Hz can be seen clearly in the projection as well as in the measured power noise. Our calculations suggest that the 5\% of cavity transmission RMS power noise that is unnaccounted for by the projections sets the maximum DC frequency offset between the cavity resonance and the laser at 7.5\,Hz.

As Figure \ref{fig:RIN} shows the spectrum of the projected pointing noise is significantly below the measured power noise spectrum in transmission of the cavity. The automatic alignment system suppresses the transmitted  power noise in two ways. First, by compensating for relative pointing fluctuations between the Eigenmode of the cavity and the input beam and second by reducing the coupling of pointing noise to transmitted power noise by maintaining the ideal alignment between the input beam and the cavity Eigenmode. Since only 5\% of the current RMS of the measured cavity power noise is unaccounted for we assume that any offset in $|\epsilon|$ must be less than 0.02.

\section{Conclusions and implications for ALPS\,II}

\label{sec:CON}

These results demonstrate that in the given environment we can stabilize a laser to a long baseline cavity with a finesse above 3000 for extended periods of time while coupling more than 95\% of the incident light into the cavity. The frequency stabilization scheme, automatic-alignment controls, and input power stabilization all work together to suppress the RMS power noise of the light in transmission of the cavity. These systems were able to reduce the RMS to 0.04\% of the maximum power transmitted through the cavity.  The current measured losses would limit the power buildup factor for the ALPS\,II production cavity. Therefore, further work must be done to characterize and mitigate the extra losses in the cavity. 

We assume that the pointing noise is predominately from the input beam and we do not believe it will significantly change for ALPS\,II. Therefore, this noise is not expected to prevent the production cavity from maintaining its required circulating power. The larger Eigenmode of ALPS\,II will actually suppress the impact of pointing noise from the input optics on the real part of $\epsilon$. Using a cavity with a larger mode, however also means that the divergence angle will be smaller and therefore the input pointing noise will have a larger effect on the imaginary part of $\epsilon$. Still the divergence angle will only decrease by a factor of 3.5 so the automatic alignment system should be able to reduce this noise enough to ensure that we can sustain the necessary circulating power in the production cavity.

We can also estimate the impact of displacement noise in the production cavity of ALPS\,II with the projected differential frequency noise derived in Section \ref{sec:RIN}. The ALPS\,II production cavity is designed to have a smaller linewidth compared to the 20\,m cavity. This is due to the fact that it is longer and will use a highly reflective end mirror in a non-impedance matched configuration. 

As Equation \ref{equ:SEI} shows, changes in the cavities resonant frequency due to seismic noise will scale inversely with the length of the cavity just as the cavity linewidth does. Therefore the coupling from displacement noise to power noise in transmission of the cavity is unaffected by changes in the cavity length. Furthermore, seismic noise measurements reveal a similar environment for the 20\,m cavity lab and the ALPS\,II site. While the differential displacement noise in ALPS\,II will have less common mode rejection due to the longer cavity length, this will occur at low enough frequencies that the frequency stabilization control loop should be able to compensate for it. 

Using a highly reflective end mirror will also lead to some reduction in the linewidth that will increase the coupling of displacement noise to the transmitted power noise. Still, we do not anticipate that a change in the cavity finesse by a factor of 2.7 will be substantial enough to prevent the cavity from meeting its requirements on circulating power.

The coupling between frequency noise in the NPRO and the transmitted power noise will increase for a longer cavity. Despite this we expect that the passive filter property of the cavity should ensure that transmitted power noise still meets its requirements. This indicates that it should be possible to maintain a stable resonance condition for the ALPS\,II production cavity. 


\vspace{.5cm}
\noindent The authors would like to thank the other members of the ALPS collaboration for valuable discussions and support and here especially Axel Knabbe and Dieter Trines for their assistance with the vacuum system as well as Reza Hodajerdi and Marian D\"urbeck for their work related to the laser system and input optics. This work would not have been possible without the wealth of expertise and hands-on support of the technical infrastructure groups at DESY. We are grateful to the DFG and particularly to the SFB 676 for funding support.

\end{document}

%% file: figures/fullFINALsmall.tex
\begingroup%
  \makeatletter%
  \providecommand\color[2][]{%
    \errmessage{(Inkscape) Color is used for the text in Inkscape, but the package 'color.sty' is not loaded}%
    \renewcommand\color[2][]{}%
  }%
  \providecommand\transparent[1]{%
    \errmessage{(Inkscape) Transparency is used (non-zero) for the text in Inkscape, but the package 'transparent.sty' is not loaded}%
    \renewcommand\transparent[1]{}%
  }%
  \providecommand\rotatebox[2]{#2}%
  \ifx\svgwidth\undefined%
    \setlength{\unitlength}{354.33070866bp}%
    \ifx\svgscale\undefined%
      \relax%
    \else%
      \setlength{\unitlength}{\unitlength * \real{\svgscale}}%
    \fi%
  \else%
    \setlength{\unitlength}{\svgwidth}%
  \fi%
  \global\let\svgwidth\undefined%
  \global\let\svgscale\undefined%
  \makeatother%
  \begin{picture}(1,0.432)%
    \put(0,0){\includegraphics[width=\unitlength]{figures/fullFINALsmall.eps}}%
    \put(0.01449547,0.3884262){\color[rgb]{0,0,0}\makebox(0,0)[lb]{\smash{Table 1}}}%
    \put(0.90464377,0.3913453){\color[rgb]{0,0,0}\makebox(0,0)[lb]{\smash{Table 2}}}%
    \put(0.21904579,0.21750923){\color[rgb]{0,0,0}\makebox(0,0)[lb]{\smash{\small AOM}}}%
    \put(0.14519993,0.14957195){\color[rgb]{0,0,0}\makebox(0,0)[lb]{\smash{\small EOM}}}%
    \put(0.55040267,0.23528198){\color[rgb]{0,0,0}\makebox(0,0)[b]{\smash{\small PZT$_1$}}}%
    \put(0.70803148,0.03143466){\color[rgb]{0,0,0}\makebox(0,0)[lb]{\smash{\small PD$_{\rm o}$}}}%
    \put(0.94905622,0.10515434){\color[rgb]{0,0,0}\makebox(0,0)[lb]{\smash{\small PD$_{\rm t}$}}}%
    \put(0.05012748,0.14088037){\color[rgb]{0,0,0}\makebox(0,0)[lb]{\smash{\small NPRO}}}%
    \put(0.285677199,0.14053015){\color[rgb]{0,0,0}\makebox(0,0)[lb]{\smash{\small Amplifier}}}%
    \put(0.55412547,0.31851494){\color[rgb]{0,0,0}\makebox(0,0)[lb]{\smash{\small QPD$_1$}}}%
    \put(0.71050000,0.31979004){\color[rgb]{0,0,0}\makebox(0,0)[lb]{\smash{\small  QPD$_2$}}}%
    \put(0.55378105,0.09201231){\color[rgb]{0,0,0}\makebox(0,0)[b]{\smash{\small PZT$_2$}}}%
    \put(0.80091883,0.04370092){\color[rgb]{0,0,0}\makebox(0,0)[lb]{\smash{\large 20\,m Cavity}}}%
    \put(0.26075795,0.07105173){\color[rgb]{0,0,0}\makebox(0,0)[lb]{\smash{\small Power stabilization}}}%
    \put(0.26171937,0.04281366){\color[rgb]{0,0,0}\makebox(0,0)[lb]{\smash{\small control electronics}}}%
    \put(0.25354908,0.36794566){\color[rgb]{0,0,0}\makebox(0,0)[lb]{\smash{\small }}}%
    \put(0.24554908,0.35133677){\color[rgb]{0,0,0}\makebox(0,0)[lb]{\smash{\small Frequency \& alignment}}}%
    \put(0.27054908,0.31972788){\color[rgb]{0,0,0}\makebox(0,0)[lb]{\smash{\small control electronics}}}%
    \put(0.62876005,0.06824435){\color[rgb]{0,0,0}\makebox(0,0)[lb]{\smash{\small PD$_{\rm i}$}}}%
    \put(0.390,0.24006507){\color[rgb]{0,0,0}\makebox(0,0)[lb]{\smash{\small Faraday}}}%
    \put(0.392,0.21629459){\color[rgb]{0,0,0}\makebox(0,0)[lb]{\smash{\small Isolator}}}%
  \end{picture}%
\endgroup%